\documentclass[aps,prl,showpacs,twocolumn,superscriptaddress]{revtex4-1}

\usepackage{amsmath,amssymb}
\usepackage{graphicx}
\usepackage{psfrag,color}

\newcommand{\xx}[1]{#1}
\newcommand{\xold}[1]{}

\newcommand{\dg}{\dot\gamma}
\newcommand{\phic}{\phi_c}
\newcommand{\deltav}{\delta\vec v}

\newcommand{\fig}[1]{Fig.~\ref{#1}}

\begin{document}

\title{Shear flow of non-Brownian suspensions close to jamming}

\author{Bruno Andreotti} \affiliation{Physique et M\'ecanique des Milieux
 H\'et\'erog\`enes, UMR 7636 ESPCI -CNRS, Univ. Paris-Diderot, 10 rue
 Vauquelin, 75005, Paris} 
\author{Jean-Louis Barrat} 
\affiliation{LIPHY, Univ. Grenoble 1; UMR 5588 et CNRS, F-38402 Saint Martin d~H\'eres, France} 
\author{Claus Heussinger}
\affiliation{Institute for Theoretical Physics, Georg-August University of
 G\"ottingen, Friedrich-Hund Platz 1, 37077 G\"ottingen,
 Germany}\affiliation{Max Planck Institute for Dynamics and Self-Organization,
 Am Fa\ss berg 17, 37077 G\"ottingen, Germay}

\begin{abstract}
  The dynamical mechanisms controlling the rheology of dense suspensions close
  to jamming are investigated numerically, using simplified models for the
  relevant dissipative forces.  We show that the velocity fluctuations control
  the dissipation rate and therefore the effective viscosity of the suspension.
  These fluctuations are similar in quasi-static simulations and for finite
  strain rate calculations with various damping schemes.  We conclude that the
  statistical properties of grain trajectories --~in particular the critical
  exponent of velocity fluctuations with respect to volume fraction $\phi$~--
  only weakly depend on the dissipation mechanism.  Rather they are determined
  by steric effects, which are the main driving forces in the quasistatic
  simulations. The critical exponent of the suspension viscosity with respect to
  $\phi$ can then be deduced, and is consistent with experimental data.
\end{abstract}

 \date{\today}

%66.20.Cy Viscosity of liquids; diffusive momentum transport, Theory and modeling
%of viscosity and rheological properties, including computer simulation

%83.80.Hj Material type, Suspensions, dispersions, pastes, slurries, colloids

\maketitle
Athermal disordered systems such as foams \cite{BW90}, emulsions \cite{CCSB09},
non-Brownian suspensions \cite{BGP11} or granular materials \cite{H10} exhibit a
critical phase transition between a liquid-like and a solid-like mechanical
behaviour, when the particle volume fraction $\phi$ crosses the jamming point
$\phi_c$. For $\phi>\phi_c$, these amorphous systems can resist shear. The
elastic shear modulus vanishes at $\phi_c$ with a critical exponent different
from the mean field one \cite{MGJS99,MW05}. Above a yield stress $\sigma_Y$,
vanishing at $\phi_c$, they present a non-Newtonian rheology, for which several
different interpretations have been proposed, based on (i) an analogy with the
glassy dynamics of a system presenting scale-free energy distributions
\cite{HL98}, (ii) interacting plastic events \cite{LC09}, (iii) the critical
scaling laws of the shear modulus and of the coordination number \cite{TWRSH10}.
Together with conventional molecular dynamics simulations (MD), quasistatic
methods (QS) have been applied to study the plastic flow of athermal amorphous
solids at the yield-stress
$\sigma_Y$~\cite{ML99,ML06,TLB06,heussingerPRL2009,HCB10}. It is generally
assumed that QS accurately describe the dynamics of the true system in the limit
of asymptotically small shear rate $\dg$. However, the existence of a proper
quasistatic limit remains controversial, and there is growing evidence that
quasistatic flows actually correspond to a finite-size dominated regime, with a
correlation length that saturates at the system size~\cite{HCB10,LC09}.
\begin{figure}[t!]
\begin{center}
\includegraphics{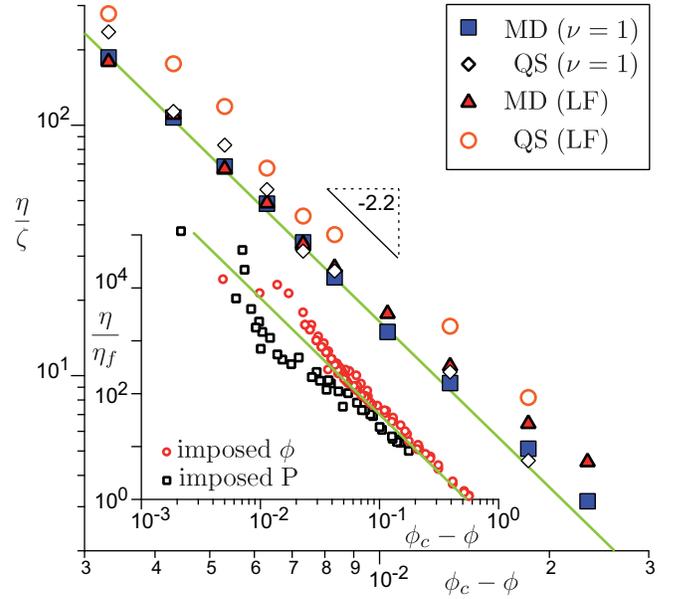}
\vspace{-4 mm}
\caption{(Color online) Viscosity $\eta$ \xx{normalized by the friction
    coefficient $\zeta$} as a function of the volume fraction $\phi$, measured
  from molecular dynamics (MD) and quasi static simulations (QS), for a
  dissipation induced by viscous drag forces of Eq. \ref{eq:friction} (labelled
  $\nu=1$) or by the lubrication like mechanism of Eq. \ref{eq:lubrication}
  (labelled LF). \xx{In MD, the viscosity is measured in the low shear rate
    regime, for $\dot \gamma=10^{-6}$ and $\zeta=10^{-1}$. In this regime,
    $\eta/\zeta$ does not depend on the precise value of these parameters, as
    shown in Fig.~\ref{fig:flowcurve}a. The quantitative agreement between
    MD($\nu=1$) and MD(LF) is coincidental.}  Inset: compilation of experimental
  data available in the literature at imposed pressure $P$ (with $\phi_c=0.587$)
  and imposed volume fraction $\phi$ (with $\phi_c=0.615$) for the ratio of
  suspension to solvent viscosity.}
\label{fig:viscosity}
\vspace{-4 mm}
\end{center}
\end{figure}

Symmetrically, for $\phi<\phi_c$, amorphous materials can flow under an
infinitesimal shear stress $\sigma$ and present a viscosity $\eta$ diverging at
$\phi_c$ like $\eta\propto (\phi_c-\phi)^{-\alpha}$.  Scaling laws are expected
to be different in thermal (glassy) systems and in athermal
sytems~\cite{IkedaBerthierSollich2012}. In the case of a suspension of
non-Brownian particles, the best fit of recent experimental results give a
critical exponent of $\alpha=2.4$ for volume\xx{-controlled} experiments
\cite{BDCL10} and of $\alpha=1.9$ for pressure\xx{-controlled} experiments
\cite{BGP11}. The explanation of the critical exponent as well as the underlying
mechanisms of the flow arrest have remained open and controversial questions up
to now.  \xx{Among the proposed mechanisms are hydrodynamic dissipation in the lubricated
films separating particles, or friction-induced normal stresses~\cite{MS09}.}
%\xx{CURRENT: For example, a mean field argument based on the average gap between
%particles, and which assumes that dissipation mostly takes place in the
%lubricated film separating particles, predicts a hydrodynamic contribution with
%an exponent $\alpha=1$~\cite{MS09}. }
A completely different interpretation
relates the divergence of the viscosity to a singular mode of the network of
contacts close to the isostatic point \cite{LernerDuringWyart2012}.
 
Here, we present simulation results for the viscous flow of a simplified model
system in the vicinity of the close-packed state at $\phi_c$. We identify a {\it
  dynamical contribution} to the divergence of the viscosity, which has its
origin in the singularity of velocity fluctuations. By comparing different
computational model systems we show that some of the statistical properties of
these velocity fluctuations are surprisingly model independent. We show how the
rheological properties, in particular the form of the flow curve and the
divergence of the viscosity can be obtained from one set of trajectories that is
based on a quasistatic simulation method.

%{\it Numerical set-up~--~}
{\it Simulations~--~} We consider a two-dimensional system constituted by $N$
soft spherical particles of mass $m$, $N/2$ of diameter $d$ and $N/2$ of
diameter $1.4d$. The particle volume fraction is defined as $\phi = \sum_{i=1}^N
\pi r_i^2/L^2$, where $L$ is the size of the simulation box. Periodic
(\xx{Lees-Edwards}) boundary conditions are used in both directions. Two
particles $i,j$ interact when their distance $r$ is smaller than the sum of
their radii $r_i+r_j$, with a repulsive potential $E(r) =
\epsilon(1-r/(r_i+r_j))^2$. All observables below are given in units of $d,m$
and $\epsilon$. We compare the divergence of the viscosity for
$\phi<\phic=0.843$~\cite{heussingerPRL2009,VagbergPRE2011Rapid} using two
different dynamics: non-equilibrium dissipative molecular dynamics (MD) and
quasistatic simulations (QS).

\begin{figure*}
\begin{center}
\includegraphics{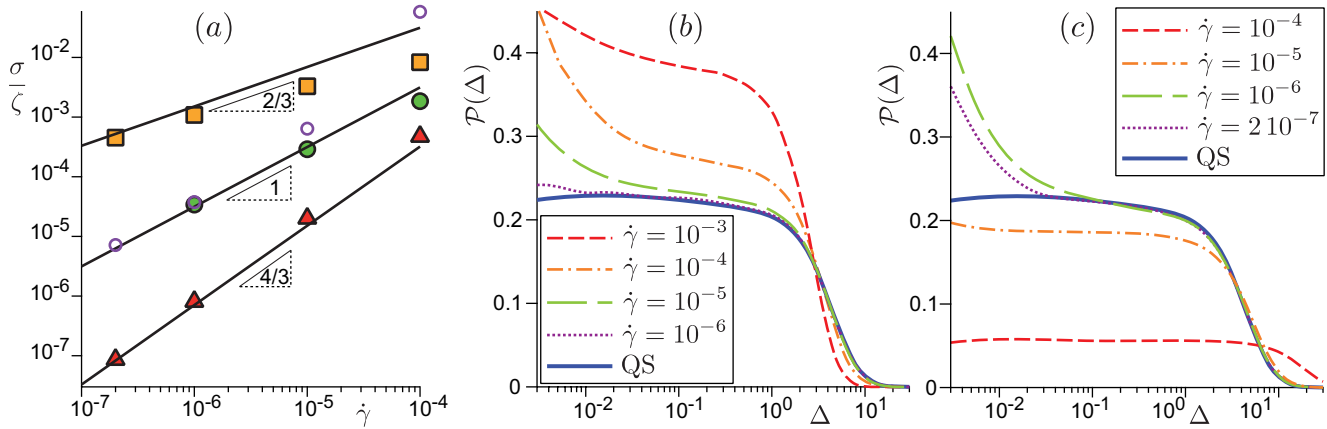}
\vspace{-4 mm}
\caption{(color online) Rheology and velocity fluctuations at $\phi=0.836$. (a)
  Relation between stress $\sigma$ and strain rate $\dot\gamma$ obtained from MD
  simulations for $\zeta=10^{-1}$, at $\nu=2/3$ ($\blacksquare$), $\nu=1$
  ($\bullet$) and $\nu=4/3$ ($\blacktriangle$), and for $\zeta=10^{-3}$ at
  $\nu=1$ ($\circ$). \xx{A small strainrate regime can be identified where
  $\sigma=\eta \dg^\nu$. The solid lines correspond to this expression, with
  $\eta$ \xx{independently} determined from the QS simulation using
  Eq.~\ref{etavanhove}. (b-c) Probability distribution function $P(\Delta)$
    of the rescaled velocity fluctuations $\Delta=\delta v/\dot \gamma$ at
    $\nu=1$, for (b) $\zeta=10^{-1}$ 
%i.e. for moderately soft particles, 
and (c)   $\zeta=10^{-3}$.}}
%, which corresponds to more rigid particles}.}
 \label{fig:flowcurve}
\vspace{-4 mm}
 \end{center}
\end{figure*}

% {\it MD simulations~--~}
In the MD simulations, the system is sheared at a shear rate $\dg$. Newton's
equations of motion $m\vec{\ddot r}_i = F_i^{\rm el}+F_i^{\rm visc}$ are
integrated with elastic contact forces $\vec F^{\rm el}=-\vec\nabla E$ and a
viscous drag force
\begin{equation}\label{eq:friction} {\vec F}^{\rm visc}({\vec v}_i) = - \zeta
  \deltav_i\left|\deltav_i\right|^{\nu-1},
\end{equation}
proportional to the $\nu$th power of the velocity difference $\deltav_i=\vec
v_i-\vec v_{\rm flow}$ between the particle velocity ${\vec v}_i$ and the flow
velocity ${\vec v_{\rm flow}(\vec r_i)}=\vec e_x \dg y$, whose fluctuations are
neglected~\cite{durianPRL1995,scalaJCP2007,olssonPRL2007,landerEPL2010}. The
flow can be viewed as being set up by a non-Newtonian fluid characterized by a
friction coefficient $\zeta$. In the special case $\nu=1$, the fluid is
Newtonian and $\zeta$ is proportional to the bare fluid viscosity $\eta_f$ (in
Stokes approximation, $\zeta= 3\pi\eta_fd$). Thermal and lubrication forces are
ignored. Unlike in granular systems, the particle-particle collisions are
elastic and the only dissipation is due to viscous losses associated with the
fluctuations of the particle velocity field.  The shear stress $\sigma$ is
calculated from the particle positions $\vec r_i=(x_i,y_i)$ and the forces $\vec
F_i=(F_{ix},F_{iy})$ acting on them as $\sigma =L^{-2}\sum_{i=1}^N x_{i}F_{iy}$.
The dominant contribution comes from the elastic forces that result from
particle overlaps. The resulting relation between the shear stress $\sigma$ and
the shear rate $\dg$ is shown in \fig{fig:flowcurve}a. For small strain rates,
both inertia and deformation of the particles are negligible, and the stress
grows with strain rate as $\sigma=\eta\dg^{\nu}$, characteristic for a power-law
fluid. \xx{The ``effective viscosity'' $\eta(\phi)$ is measured in this regime
  and is a function of the volume-fraction, as shown in Fig.~\ref{fig:viscosity}
  for the Newtonian case ($\nu=1$). At larger strainrates and in weakly damped
  systems ($\zeta=0.001$ or $\nu>1$) one observes a shear thickening regime,
  which can be ascribed to inertia.  Conversly, for stronger damping
  ($\zeta=0.1$ and $\nu<1$) and, in particular for volume fractions close to
  $\phi_c$ [23] one observes a shear thinning regime, when particle deformation
  starts to be relevant.}

%{\it QS simulations~--~}
Quasistatic simulations consist of successively applying small steps of shear
and minimizing the total potential energy. By construction, they generate
particle trajectories at $\dot\gamma\to 0$. An elementary strain step of
$\gamma_0=5\cdot10^{-5}$ is used. After each change in boundary conditions the
particles are moved affinely to define the starting configuration for the
minimization, which is performed using conjugate gradient
techniques~\cite{lammps}. The minimization is stopped when the nearest energy
minimum is found. As no static, force-balanced state exists below the jamming
point ($\phi<\phic$), the inter-particle forces at the minimum are strictly
zero; i.e. the particles can always arrange in such a way as to avoid mutual
overlaps. Thus, each minimized configuration corresponds to a true hard-sphere
state and the resulting particle trajectories can be viewed as a sequence of
snapshots of a flowing hard-sphere system at zero temperature. Particle motion
in such a system is driven by steric exclusion and the lack of free volume. In
particular, particles have to move over larger distances when the jamming point
is approached, to find a new overlap-free state compatible with the imposed
shear~\cite{heussingerEPL2010}.

Without particle overlaps all contact forces and therefore the shear stress are
strictly zero in the QS simulation. Still, an effective shear stress and
viscosity can be obtained from the power $\Gamma$ per unit surface that would be
dissipated along the QS trajectories, if the dissipation mechanism of
Eq.~(\ref{eq:friction}) was present. $\Gamma$ is equal to the power injected per
unit volume in the system, $\sigma\dot\gamma$, and can be expressed as:
\begin{equation}\label{eq:diss.en}
  \Gamma =L^{-2} \left\langle\sum_i \vec F^{visc}(\vec v_{i,\rm qs}) \cdot \left
      (\vec v_{i,\rm qs}-\vec v_{\rm flow}(\vec
      r_i)\right)\right\rangle\nonumber\\ 
\end{equation}
From this expression, we deduce the viscosity:
\begin{equation}
  \eta =\frac{\Gamma}{\dot\gamma^{1+\nu}}=- \zeta \frac{N}{L^2}\,\frac{\left\langle
      \delta v\,^{1+\nu}\right\rangle}{\dot\gamma^{1+\nu}}=- \zeta \frac{N}{L^2} \int \Delta^{1+\nu}
  P(\Delta)d \Delta\,.  
\label{etavanhove}
\end{equation}
where $P(\Delta)$ is the probability distribution function of the particle
velocity rescaled by the shear rate: $\Delta= \delta v/\dot\gamma$. As particle
coordinates in the QS simulation are only available at discrete steps, one has
to define an effective particle velocity $ \vec v_{\rm qs}=\dot\gamma\delta\vec
r/\gamma_0$ from the particle displacement $\delta\vec r$ during such a single
step. Therefore, $\Delta$ is also the displacement rescaled by the strain
interval $\gamma$ and $P(\Delta)$ is the van Hove function. Note that the
viscosity is related to the $(\nu+1)$st moment of the velocity fluctuations, and
thus of $P(\Delta)$ (Eq.~\ref{etavanhove}).

%{\it Statistical equivalence of MD and QS trajectories~--~} 
{\it Results~--~} To characterize statistically the trajectories we consider the
probability distribution for particle velocities, $P(\Delta)$. We concentrate on
the velocity component in the gradient direction ($y$-component), which
automatically eliminates trivial particle motion due to the average flow field.
When the strain rate is small enough, $P(\Delta)$ reaches a limiting form
(dotted line in Fig.~\ref{fig:flowcurve}b-c), which is directly related to the
small strain-rate power-law regime of Fig.~\ref{fig:flowcurve}a. \xx{Whenever
  the rheology $\sigma(\dot \gamma)$ deviates from this asymptotic behaviour,
  the distribution function $P(\Delta)$ deviates from its asymptotic form as
  well. Interestingly, the approach towards this asymptotic form is rather
  different in the weakly damped, i.e. shear-thickening, system
  (Fig.\ref{fig:flowcurve}b and $\bullet$ in Fig.\ref{fig:flowcurve}a) as
  compared to the strongly damped, shear-thinning system
  (Fig.~\ref{fig:flowcurve}c and $\circ$ in Fig.~\ref{fig:flowcurve}a).}

%  For weakly damped systems ($\zeta=0.001$),
%In the limit of rigid spheres ($\zeta \ll 1$), 
%  one observes a shear-thickening regime at large $\dot \gamma/\zeta$ which can
%  be ascribed to inertia ( Fig.~\ref{fig:flowcurve}c and $\circ$ in
%  Fig.~\ref{fig:flowcurve}a).  
%For softer spheres, 
%  For stronger damping ($\zeta=0.1$ and $\nu<1$) one observes conversely a shear
%  thinning regime~\cite{olssonPRL2007}, when particle interpenetration starts to
%  be relevant
  % when the typical interpenetration $\sim \sigma$ becomes comparable to the
  % typical gap $\sim (\phi_c-\phi)$ separating grains
%  (Fig.~\ref{fig:flowcurve}b and coloured symbols in Fig.~\ref{fig:flowcurve}a).
%}

\begin{figure}
\begin{center}
\includegraphics{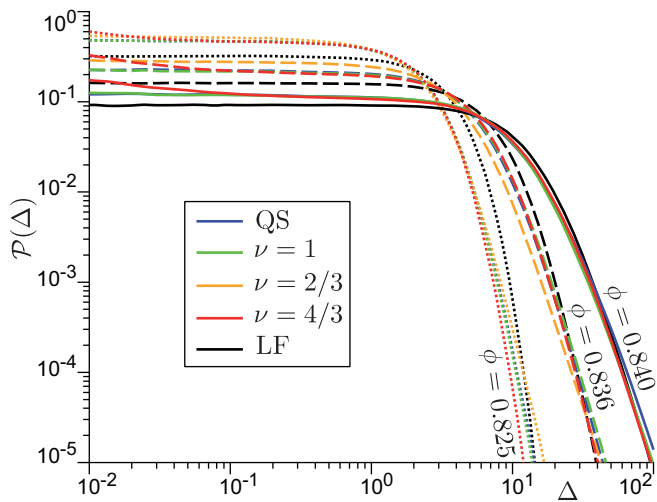}
\vspace{-4 mm}
\caption{\xx{Comparison of the probability distribution function $P(\Delta)$ of
    the rescaled velocity fluctuations $\Delta=(v-v_{\rm flow})/\dg$ obtained
    for the different computational models at different volume fractions $\phi$.
    Measurements are performed in the low shear rate asymptotic regime, for
    $\dg=10^{-6}$ and $\zeta=0.1$}.  The three values of $\nu$ correspond to the
  dissipation mechanism of Eq.  \ref{eq:friction}, the LF label refers to the
  lubrication like mechanism of Eq. \ref{eq:lubrication} \xx{and QS, to
    quasi-static simulations}.  }
 \label{fig:vh.diff.models}
\vspace{-4 mm}
 \end{center}
\end{figure}

\xx{The velocity fluctuation PDFs obtained in the QS simulation (solid lines in
  Fig~\ref{fig:flowcurve}b-c) are similar to those obtained in MD in the limit
  of vanishing $\dot \gamma$.  }In particular, the sharp shoulder at
$\Delta\approx 4$ is well reproduced for both strongly ($\zeta=10^{-1}$) and
weakly ($\zeta=10^{-3}$) damped systems. \xx{Furthermore, as shown in
  Fig.~\ref{fig:vh.diff.models}, the small strain-rate form of $P(\Delta)$ only
  weakly }depends on the value of the exponent $\nu$. Again, the most pronounced
feature is the shoulder, which is nearly identical in all four simulations.
However, small differences between MD and QS remain especially for small damping
(Fig.~\ref{fig:flowcurve}c) or $\nu>1$ (Fig.~\ref{fig:vh.diff.models}). Here, a
rise at small $\Delta\to 0$ is observed at the smallest strainrates, which is
not present in the QS simulation. \xx{Importantly, such small velocities do not
  contribute to the second moment of the distribution and therefore are
  irrelevant for the dissipated energy and the viscosity (Eq.~\ref{eq:diss.en}).
  By way of contrast, these small differences may be important for the number of
  inter-particle contacts. As it turns out, the coordination number $Z$, which
  is the number of contacts per particle (taken from simulation snapshots),
  strongly varies with either $\nu$ or $\zeta$ and is also different in the QS
  simulation;
%Interestingly, the inter-particle connectivity does not display a similar
%small-strainrate regime. The connectivity, as measured by the average
%instantaneous number of overlaps per particle, strongly varies with either $\nu$
%or $\zeta$ and is also different in the QS simulation; 
its value changes from $Z_{\rm iso}\lessapprox 4$, i.e. close to the isostatic
state, down to small values $Z<1$.}
%  Thus, in our system the value of $z$ has no
%immediate predictability and a more refined definition of contacts is
%necessary.}

We conclude that the overall features of the particle trajectories in the MD
simulations are statistically comparable to those in the QS simulation. The
small strain-rate power-law fluid regime (Newtonian for $\nu=1$) should
therefore be considered as a true quasi-static limit, which is by no means
obvious. In fact, the QS limit seems much better defined here ($\phi<\phic$)
than in the plastic flow regime ($\phi>\phic$), where QS simulations have
usually been applied, but where they suffer from a dependence on system
size~\cite{LC09,HCB10}.

%{\it Viscosity exponent close to jamming~--~} 
One important consequence of the equivalence MD-QS is that one set of QS
trajectories can be used to determine the flow rheology for different values of
$\nu$.  Fig.~\ref{fig:flowcurve}a compares the rheology obtained using MD (data
points) and QS simulations (solid line). They nicely collapse on each other when
MD simulations are considered in the limit of small shear rate.
Fig.\ref{fig:viscosity} shows the viscosity $\eta$ determined from both
simulations, as a function of volume fraction $\phi$. Beyond noting the quality
of the collapse, one observes that the viscosity diverges with $\phi_c-\phi$,
with a scaling exponent $\simeq 2.2$ consistent with the values measured
experimentally.

Eq.(\ref{etavanhove}), giving the power dissipated per unit volume, leads to the
scaling law $\eta\sim \langle \delta v^{1+\nu}\rangle $, which connects the
divergence of the macroscopic viscosity to the scaling law followed by the
microscopic particle motion $\delta v\sim (\phi_c-\phi)^{-\beta}$.  Thus, the
seemingly harmless power balance turns into a relation between the exponents
controlling the divergence of velocity fluctuations and that of viscosity:
$\alpha=\beta(1+\nu)$. We have recently shown that $\beta\approx
1.1$~\cite{heussingerEPL2010}, which gives (for $\nu=1$) $\alpha\simeq 2.2$,
consistent with the exponent extracted from the MD data in
Fig.~\ref{fig:viscosity}. Note however, that subdominant corrections can lead to
apparent exponents that, in the considered range of densities, may not reflect
the true asymptotic
behavior\cite{olssonPRE2011scaling.corrections,VagbergPRE2011Rapid}.

{\it Discussion~--~} The small strain-rate rheology, $\sigma=\eta\dg^\nu$, as
well as the divergence of the viscosity, $\eta\sim\delta\phi^{\beta(1+\nu)}$,
depend on the value of $\nu$. On the other hand, the underlying particle
trajectories are hardly affected by changing $\nu$.  This points to a certain
decoupling between particle trajectories and dissipative process. In this
picture, the statistical properties of trajectories are largely governed by the
structural singularity of random close packing and the lack of space available
for particle motion. On the other hand, system-specific dissipation mechanisms
affect the rheological properties via the dissipated energy along these
geometrically predetermined trajectories. Certainly, such a decoupling cannot be
realized in a perfect manner, as shown by the small differences of $P(\Delta)$
at small $\Delta$ and in the tails (Fig.~\ref{fig:flowcurve} and
Fig.~\ref{fig:vh.diff.models}).  Nevertheless, it seems to be strong enough such
that the various flow curves (Fig.~\ref{fig:flowcurve}a) and the viscosity
(Fig.~\ref{fig:viscosity}) can accurately be predicted from the sole knowledge
of one set of QS trajectories. \xx{The scaling law relating the viscosity to the
  volume fraction is a directly testable prediction of the central idea of this
  letter. It suggests to measure the rheology of particles suspended in a
  non-Newtonian solvent like a polymer melt or a visco-plastic fluid.}

In order to investigate the universality of the decoupling phenomenon, we have
conducted additional simulations with a damping that describes a modified
lubrication force~\cite{footnote} between neighboring particles $i$ and $j$:
\begin{eqnarray}\label{eq:lubrication}
\vec F^{\rm diss}(\vec v_i,\vec v_j) = -\zeta\hat n_{ij}\left[\hat n_{ij}\cdot
 \left(\vec v_i-\vec v_j\right)\right]\,,
\end{eqnarray}
In agreement with previous (overdamped)
simulations~\cite{olssonPRL2007,TWRSH10}, the spatial velocity-correlation
function is qualitatively different in the models of Eqs.(\ref{eq:friction}) and
(\ref{eq:lubrication}) (not shown).  Still, the probability distribution
$P(\Delta)$ (Fig.\ref{fig:vh.diff.models}) is remarkably similar across all
models considered. In particular, the scale of velocity fluctuations increases
on approaching the critical volume-fraction $\phi_c$, and the overall agreement
between the different curves seems to improve.  This supports our interpretation
of the role of close packing for the particle trajectories. Furthermore,
starting with the QS trajectories, a calculation similar to
Eq.~(\ref{eq:diss.en}) can again be used to predict the viscosity. As
Fig.\ref{fig:viscosity} illustrates (open circles) this calculation is quite
accurate but slightly overestimates the true viscosity (triangles), roughly by a
factor $\approx 1.5$.

In conclusion, singular velocity fluctuations cause a dynamical contribution to
the divergence of the viscosity, \xx{as independently noted in
Ref.~\cite{LernerDuringWyart2012}}. These velocity fluctuations are surprisingly
conserved across different computational models, which we explain with geometric
features and the lack of available space close to the jamming transition.
\xx{Our results complement those obtained in Ref.~\cite{LernerDuringWyart2012},
  for frictionless hard spheres. In that system, the jamming transition
  coincides with the isostatic threshold and the flow properties can be related
  to the geometry of the contact network. The coordination number $Z$ is then
  the relevant control parameter. Our results bridge the gap between such an
  ideal system and those where inertia and elasticity lead, for a given volume
  fraction $\phi$, to strong changes of $Z$.
  %As it turns out, the
%  coordination number, taken from simulation snapshots, strongly depends in MD
%  on $\nu$ and $\zeta$, and is also different in QS; its value changes from
%  $Z_{\rm iso}\lessapprox 4$, i.e. close to the isostatic state, down to small
%  values $Z<1$.  
  Indeed, in our simulations, the value of $Z$ has no immediate predictability
  and $\phi$ is the only relevant parameter.  Our results open the promising
  perspective of the existence of an inherent contact network which would govern
  the topography of the energy landscape.  Such a concept would establish the
  missing connection between volume fraction and connectivity.}

%Interestingly, the inter-particle connectivity does not display a similar
%small-strainrate regime. The connectivity, as measured by the average
%instantaneous number of overlaps per particle, strongly varies with either $\nu$
%or $\zeta$ and is also different in the QS simulation; 

\acknowledgments
BA and JLB are supported by Institut Universitaire de
France; CH is supported by the Deutsche Forschungsgemeinschaft, Emmy Noether
program: He 6322/1-1; JLB acknowledges a useful discussion with A. Lema\^itre.


\begin{thebibliography}{10}
\bibitem{footnote} In order to ensure Newtonian flow at small strain rates we choose an interaction range $r_c^{\rm diss}=1.5(r_i+r_j)$ that is larger than the range for elastic interactions. In effect, this corresponds to a lubrication force with a divergence that is cut off at short distances.
\end{thebibliography}
\end{document}